\documentclass[aps,prb,twocolumn,amsmath,amssymb,showpacs]{revtex4}
\usepackage{graphicx}
\usepackage{Jonasmacros}
\usepackage{bm}
\usepackage{mathptmx}

\usepackage[usenames,dvipsnames]{color}

\begin{document}
\title{Molecular graphene under the eye of scattering theory}

\author{H. Hammar}
\author{P. Berggren}
\author{J. Fransson}
\email{Jonas.Fransson@physics.uu.se}
\affiliation{Department of Physics and Astronomy, Uppsala University, Box 516, SE-751 21 UPPSALA, Sweden}

\begin{abstract}
The recent experimental observations of designer Dirac Fermions and topological phases in molecular graphene are addressed theoretically. Using scattering theory we calculate the electronic structure of finite lattices of scattering centers dual to the honeycomb lattice. In good agreement with experimental observations, we obtain a V-shaped electron density of states around the Fermi energy. By varying the lattice parameter we simulate electron and hole doping of the structure and by adding and removing scattering centers we simulate respectively vacancy and impurity defects. Specifically for the vacancy defect we verify the emergence of a sharp resonance near the Fermi energy for increasing strength of the scattering potential.
\end{abstract}
\maketitle

The interest in Dirac Fermions has grown for a wide community after the first synthesis of a monolayer graphene and the subsequent observations of massless Fermions.\cite{novoselov2004,geim2007,katsnelson2007,castroneto2009,vozmediano2010} As a common feature to a new class of materials that has emerged after the initial break through, the band structure and embedded spin degree of freedom of the Dirac Fermions is described by the relativistic Dirac equation.\cite{katsnelson2007,castroneto2009}

There has been an increasing interest in engineered systems that share key properties, including Dirac Fermions, with graphen.\cite{polini2013} The hexagonal lattice structure has shown crucial for the formation of Dirac Fermions, experimentally realized by for example confining photons in hexagonal patterns, \cite{peleg2007,kuhl2010} nanopatterning of ultra-high-mobility two-dimensional electron gase,\cite{singha2011} scanning probe methods to assemble molecules on metallic surfaces,\cite{gomes2012} and trapping ultracold atoms in optical lattices.\cite{soltan2011,tarruell2012,uehlinger2013} It is important to notice that artificial systems provide alternative routes for studies of topological \cite{haldane1988} and quantum spin Hall insulators,\cite{kane2005,guinea2010} as well as to novel non-trivial strongly correlated phases.\cite{meng2010}

Although experimental progress within the field of artificially assembled nanostructures has been tremendous, as shown in the above examples, the theoretical advances has focused less on aspects of engineered nanostructures. Here, we theoretically study molecular graphene which is constructed by depositing scattering defects in a regular triangular lattice on a metallic surface. Using scattering theory,\cite{fiete2003} we calculate the local density of electron states within the engineered lattice structure and find a linear spectrum around the Fermi level, in excellent agreement with experiments.\cite{gomes2012} Further, we study the effects of electron and hole doping by modifying the lattice parameter, and magnetic field effects by imposing strain to the lattice. Especially, we consider single defect scattering, and simulate both vacancy and impurity defects. For the vacancy defects we verify the emergence of a sharp resonance near the Fermi level for increasing strength of the vacancy scattering potential.

Previous attempts of modeling molecular, or artificial, graphene have focused on the implementation of infinite lattice structures in two-dimensional electron gases using tight-binding models.\cite{wunsch2008,park2009,gibertini2009} Here, we instead employ a scattering theoretical approach which allows us to study finite structures of arbitrary size. The flexibility this approach offers has been proven invaluable in previous studies of for example engineered nanostructures on metallic surfaces,\cite{fiete2003,fransson2010,gawronski2011} and single defects on topological insulator surface \cite{she2013} and graphene.\cite{fransson2013} Our results presented here are, hence, also an important demonstration of a tool that can be applied more generally for studies of atomic and molecular assemblies embedded in a two-dimensional electron gas.

Local probing techniques are especially promising and useful for building understanding of the interactions in low-dimensional materials and complexes of molecular structures.\cite{hirjibehedin2006,zhou2010,decker2011,wang2012} Such techniques have been employed in very different contexts for engineering assembled nanostructures aiming towards nano-magnetic memories,\cite{loth2012} spin-based logic gates,\cite{khajetoorians2011} and coherent quantum phase measurements.\cite{moon2008}

Scanning tunneling microscopy (STM) is a local probing technique \cite{binnig1982} which offers a route for mapping the local electronic structure using the tunneling current flowing between the STM tip and a surface.\cite{tersoff1983} At low temperatures the tunneling conductance is given by
\begin{align}
\frac{dI(\bfr,V)}{dV}\propto&\, 
	n_\tip(\dote{F}-eV)N(\bfr,\dote{F}),
\end{align}
where $\dote{F}$ is the Fermi level of the system in equilibrium, whereas $V$ is the source-drain voltage applied across the tunneling junction. This expression relates the tunneling (differential) conductance ($dI/dV$) to the local electronic densities of states (DOS) of the surface ($N$) and the tip ($n_\tip$). Typically, the tip electronic density is featureless so that signatures picked up in the tunneling conductance can be attributed to variations in the local surface density of electron states $N(\bfr,\omega)$. Calculating this density is henceforth our primary focus.

We consider a metallic surface modeled by a two-dimensional electron gas using $\Hamil_0=\sum_{\bfk\sigma}\dote{\bfk}\cdagger{\bfk}\cc{\bfk}$, where $\cdagger{\bfk}$ ($\cc{\bfk}$) creates (annihilates) an electron with energy $\dote{\bfk}$, momentum $\bfk$ and spin $\sigma=\up,\down$. Scattering points are inserted at the positions $\bfR_n$ through the energy $\Hamil_{\rm int}=\sum_nV_nn(\bfR_n)$, where $n(\bfr)=\sum_\sigma\int\cdagger{\bfk}\cc{\bfk'}e^{-i(\bfk-\bfk')\cdot\bfr}d\bfk d\bfk'/(2\pi)^4$ is the electronic charge at the spatial position $\bfr$. The surface electron density can be calculated through the relation $N(\bfr,\omega)=-\im G(\bfr,\bfr;\omega)/\pi$, where $G(\bfr,\bfr';\omega)$ is the surface Green function (GF) which describes the local electronic structure. The real space GF connects to the introduced model through the Fourier transform $G(\bfr,\bfr';\omega)=\int G_{\bfk\bfk'}(\omega)e^{i\bfk\cdot\bfr-i\bfk'\cdot\bfr'}d\bfk d\bfk'/(2\pi)^4$. Here, we have suppressed the spin indices since our system is assumed to be perfectly spin-degenerate.

\begin{figure}[t]
\begin{center}
\includegraphics[width=.99\columnwidth]{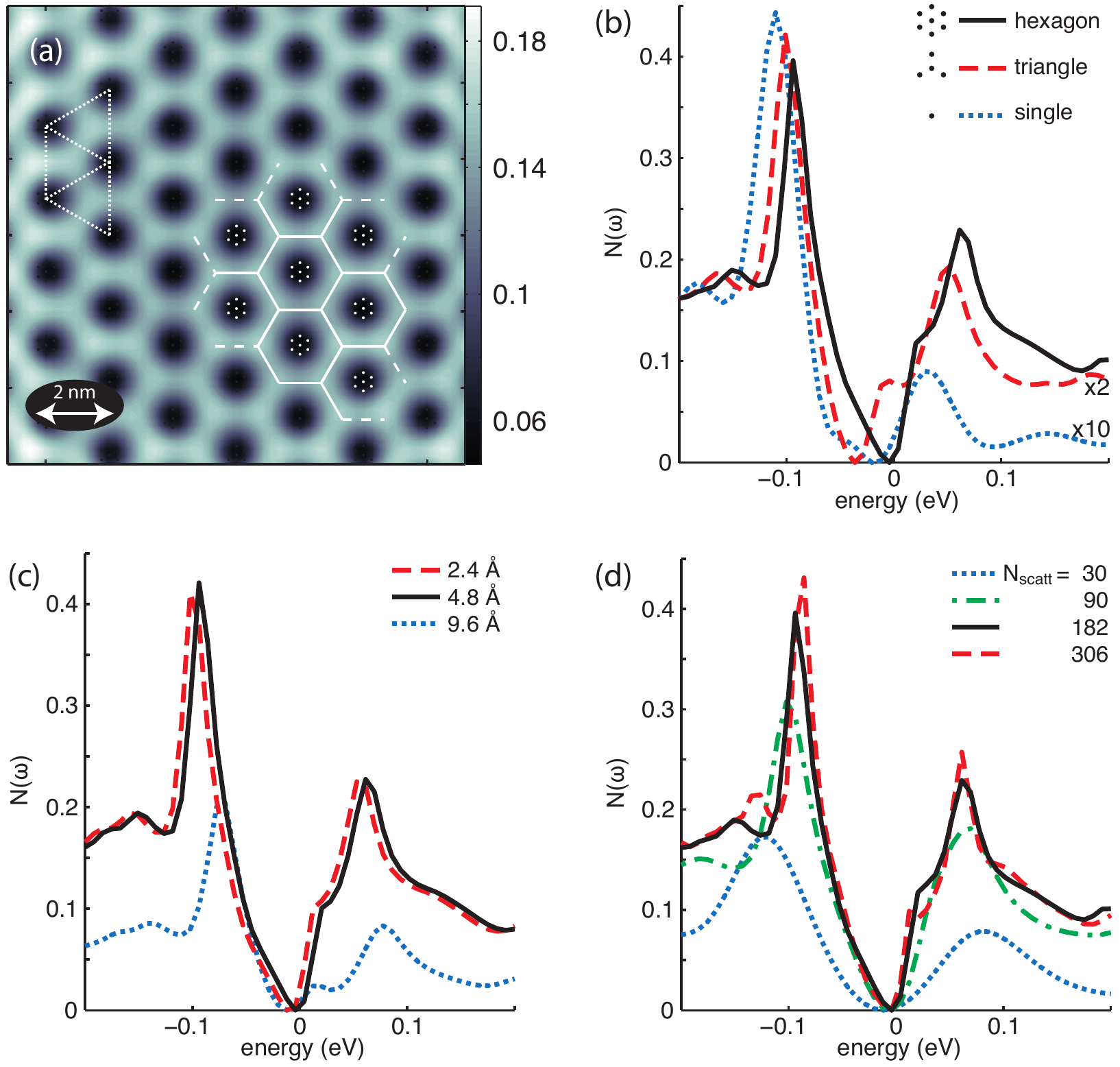}
\end{center}
\caption{(a) Topograph of a lattice consisting of hexagonal scatterers at the energy $-10$ meV. The triangular and hexagonal lattices are added as guides to the eye. (b) - (d) Local DOS as function of (b) number of scattering points in each scattering center (7 points - hexagon, 4 points - triangle, and single point), (c) diameter of the (hexagonal) scattering centers, and (d) lattice size: $5\times6$, $9\times10$, $13\times14$, and $17\times18$ (hexagonal) scattering centers. In panels (a), (b), and (d) we used the diameter $4.8$ \AA\ for the hexagonal scattering centers and lattice parameter $a=19.2$ \AA. For all plots we have used $\dote{\bfk}-\dote{F}=E_0+\hbar^2k^2/2m^*$, with $E_0\simeq-0.45$ eV pertaining for Cu(111), and $m^*=0.38m$, where $m$ is the free electron mass.}
\label{fig1}
\end{figure}

The \emph{dressed} GF $G$ is constructed through a $T$-matrix expansion, see for example \cite{fiete2003,fransson2012} for more details,
\begin{subequations}
\begin{align}
G_{\bfk\bfk'}(\omega)=&
	\delta(\bfk-\bfk')g_\bfk(\omega)
\nonumber\\&
	+\sum_{mn}
		g_\bfk(\omega)e^{-i\bfk\cdot\bfR_m}
		{\cal T}(\bfR_m,\bfR_n)
		e^{i\bfk'\cdot\bfR_n}g_{\bfk'}(\omega),
\\
{\cal T}(\bfR_m,\bfR_n)=&
	t(\bfR_m,\bfR_n)V_n,
\end{align}
\end{subequations}
where $t^{-1}(\bfR_m,\bfR_n)=\delta(\bfR_m-\bfR_n)-V_mg(\bfR_m-\bfR_n)$, $g(\bfr-\bfr')=\int g_\bfk e^{i\bfk\cdot(\bfr-\bfr')}d\bfk/(2\pi)^2$, and the \emph{bare}, or unperturbed, GF $g_\bfk=(\omega-\dote{\bfk}+i\delta)^{-1}$, where $\delta>0$ is infinitesimal. This expansion describes the influence of scattering at the positions $\bfR_m$ on the electronic structure, by providing a correlation between electron creation at $\bfr'$ and annihilation at $\bfr$ under the presence of potential scattering.

The scattering potentials at the sites $\bfR_m$ depletes the electron density in a neighborhood around those positions, electron density which has to be redistributed elsewhere in the structure. This generates a new electronic structure. By distributing the scattering potentials according to a triangular (dotted lines), we construct an electronic density which is distributed as a honeycomb lattice, see Fig. \ref{fig1} (a).

In an experimental set-up, the scattering centers would be represented by atomic or molecular entities,\cite{gomes2012} which provides a continuous and spatially extended potential landscape. While we have used Dirac point like scattering potentials, $V_mn(\bfR_m)=\int V_0n(\bfr)\delta(\bfr-\bfR_m)d\bfr$ in our calculations, spatial extension of the scattering centers can be obtained by inserting several scattering points in cluster formations, see legend of Fig. \ref{fig1} (b). The electron DOS resulting from three different types of scattering centers are plotted in Fig. \ref{fig1} (b), single scattering points (dotted), four points in triangular (dashed), and seven points in hexagonal (solid) form. The plotted electron DOS are obtained by spatially averaging around the scattering centers and subtracting the flat back ground density of the surface states. The plots clearly illustrate how the spatial extension of the scattering centers build up the linear spectrum around the Fermi level (zero energy point), the Dirac point, and we base our following discussion on the hexagonally shaped scattering centers.

\begin{figure}[t]
\begin{center}
\includegraphics[width=.99\columnwidth]{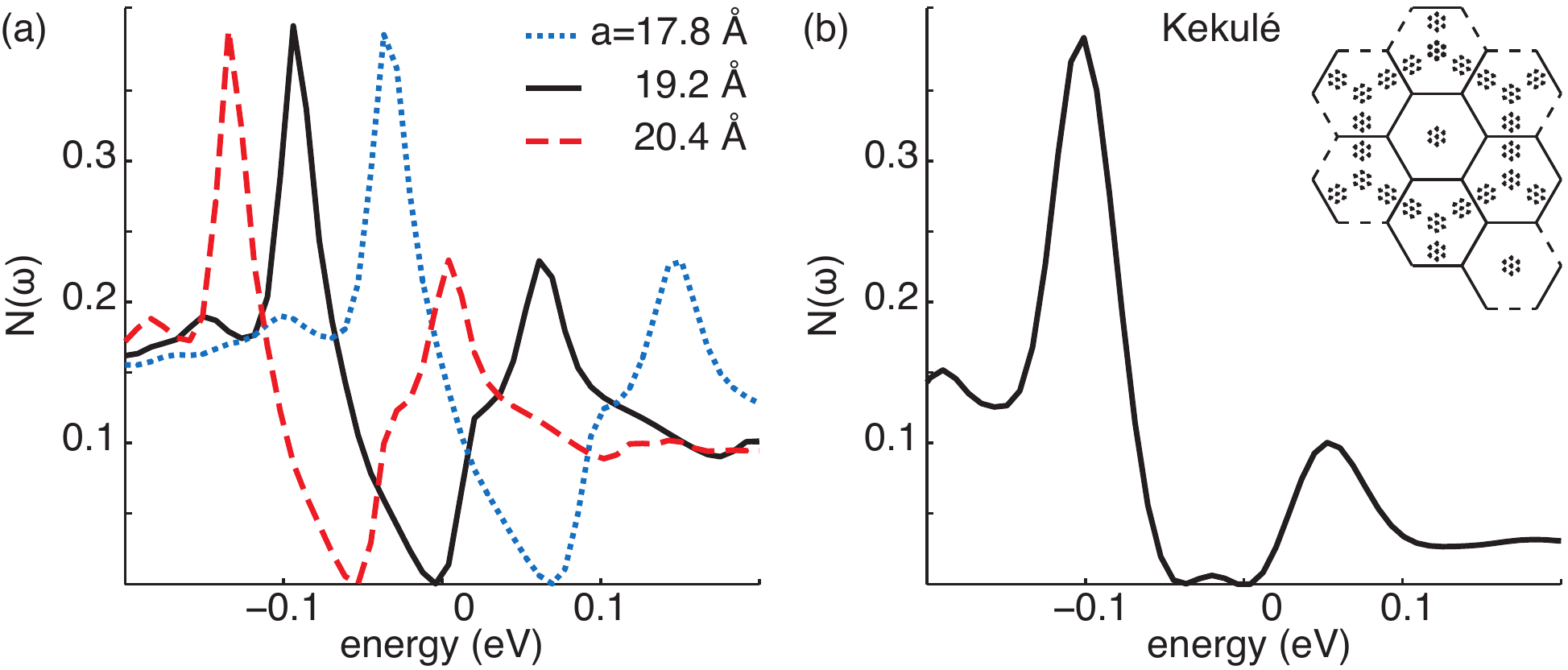}
\end{center}
\caption{Electron DOS for (a) varying lattice parameter $a$, going from hole doped (dotted) via nearly neutral (solid) to electron doped (dashed) molecular graphene, and (b) Kekul\'e texturing of the scattering centers. Inset shows the geometry of the Kekul\'e texturing. Other parameters are as in Fig. \ref{fig1}.}
\label{fig2}
\end{figure}

Furthermore, we test the importance of the spatial extension of the scattering centers by varying the diameter of the hexagonally shaped scattering centers, and in Fig. \ref{fig1} (c) we plot the electron DOS for different sizes. The resulting DOS point towards the fact that scattering centers of about 2 | 5 \AA\ in diameter, are sufficiently large to \emph{not} have a point like influence on the electronic density. Too large scattering centers (diameter $\gtrsim 9$ \AA) tend, on the other hand, to modify the electronic density towards a double well structure around the Fermi level.

The lattice size has a natural influence on the electron DOS at the center of the lattice, see Fig. \ref{fig1} (d), where we plot the electron DOS as function of the lattice size. Distributing scattering centers with a diameter of 4.8 \AA\ in a triangular lattice with lattice parameter $a=19.2$ \AA, clearly indicates that the central electron DOS converges towards a linear spectrum around the Fermi level sufficiently well for 100 | 200 scattering centers arranged in a nearly quadratic form.

Similarly, changes in the lattice parameter $a$ has the influence of doping the molecular graphene. Smaller (larger) lattice parameter functions as a stronger (weaker) confinement of the electron density, which thereby pushes the spectrum towards higher (lower) energies, see Fig. \ref{fig2} (a). In this way one can achieve electron (hole) doping of the system. Apart from the rigid shift, the shape of the spectrum remains nearly unaffected by the changes in the lattice parameter.

The flexibility offered by scattering theoretical approach is illustrated by the spectrum plotted in Fig. \ref{fig2} (b), showing the electron DOS for a Kekul\'e textured lattice, see inset. In excellent agreement with the experiments,\cite{gomes2012} we reproduce the opening of a finite gap at the Dirac points, as well as the strong peaks appearing on each side.

\begin{figure}[t]
\begin{center}
\includegraphics[width=.99\columnwidth]{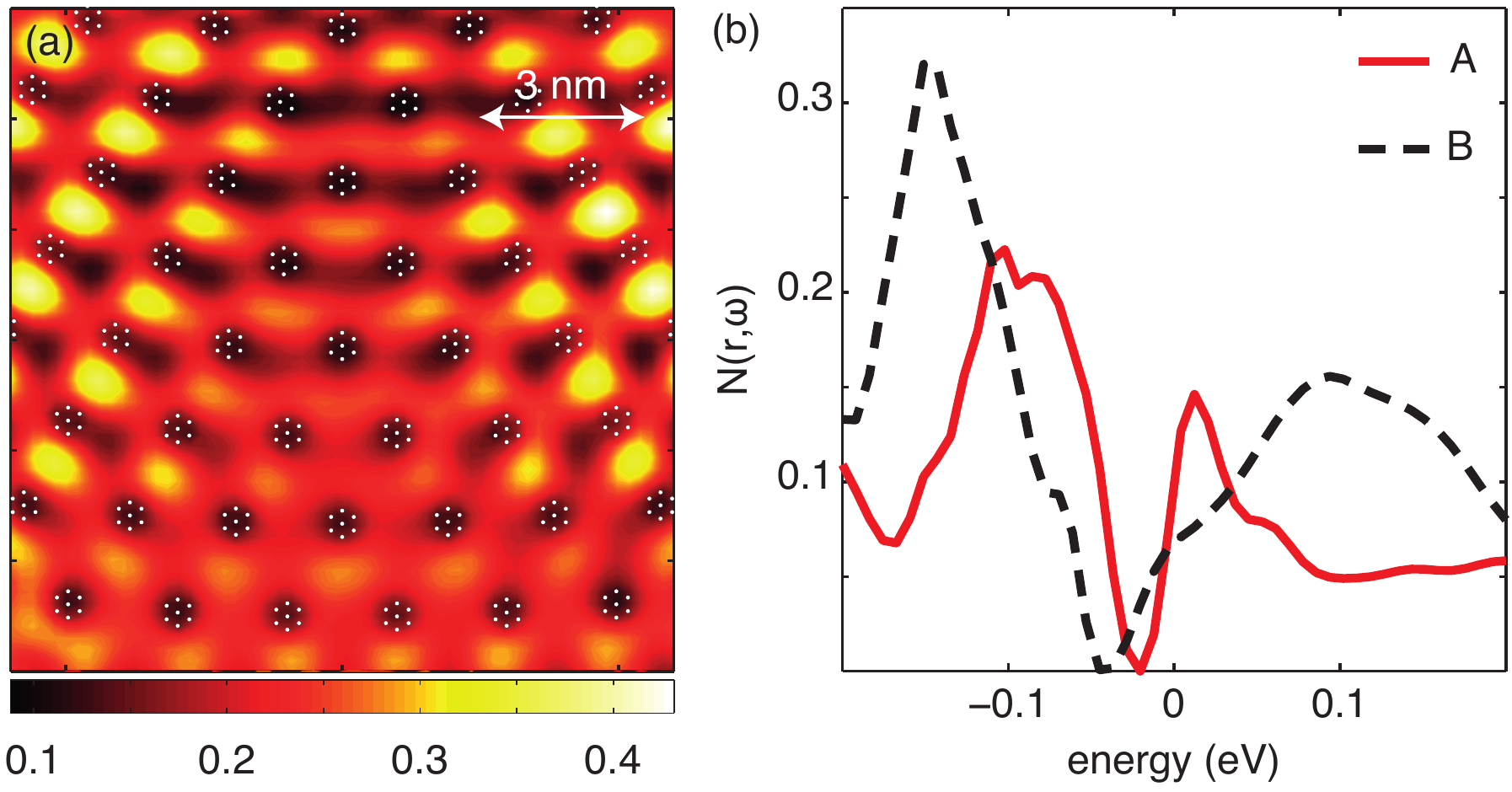}
\end{center}
\caption{Molecular graphene under a pseudo-magnetic field corresponding 60 T ($q=10^{-3}$ \AA). (a) Spectral density and (b) electron DOS in the A (solid) and B (dashed) sublattices.}
\label{fig3}
\end{figure}

Imposing strain, or pseudo-magnetic field, to molecular graphene has the same effect as in real graphene, namely breaking the pseudo-spin symmetry of the Dirac point. Practically, a constant strain field is introduced through displacement of the scattering centers, in polar coordinates $(r,\theta)$, using $(u_r,u_\theta)=(qr^2\sin3\theta,qr^2\cos3\theta)$, where $q$ is the a parameter for the strength of the strain.\cite{guinea2010}

In Fig. \ref{fig3} (a) we show the spectral density at the center of the lattice for strain conditions corresponding to a magnetic field of 60 T. In agreement with experiments, the strain breaks the pseudo-spin symmetry of the Dirac point which becomes visible in the spatial resolution of the spectral density. This results in the formation of an $A$-sublattice with increased density at the Fermi level (bright spots) and a $B$-sublattice with reduced density (dark spots). This is more clearly shown in Fig. \ref{fig3} (b) where we plot the local DOS associated with respective sublattice. The local DOS shows a well-defined zero energy state | the zero Landau level | in the bright spots of the $A$-sublattice (solid). The dark regions of the $B$-sublattice (dashed) are associated with a reduced electron density, revealing the Landau gap below the Fermi level.

\begin{figure}[t]
\begin{center}
\includegraphics[width=.99\columnwidth]{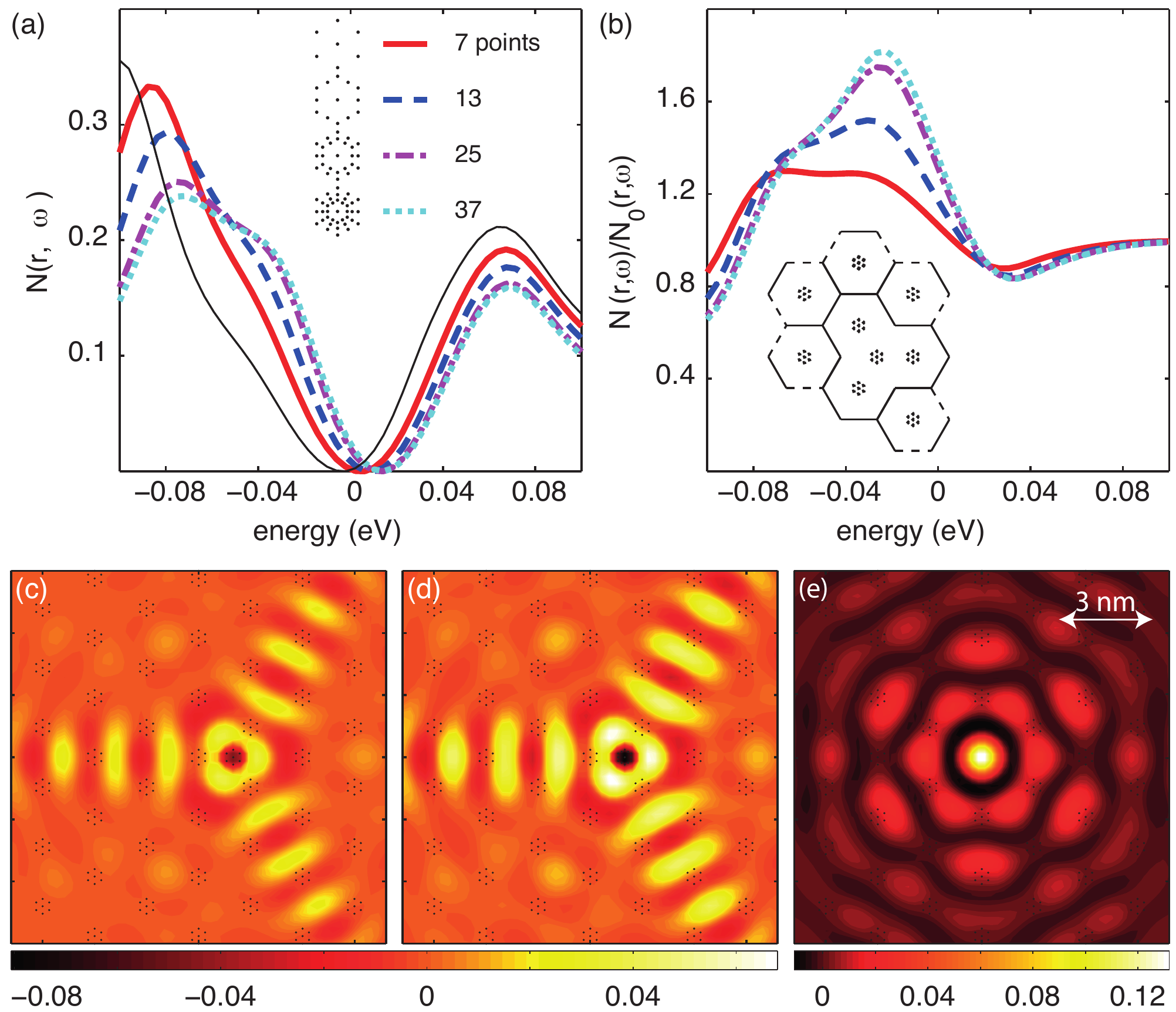}
\end{center}
\caption{Effect of single defect in molecular graphene. (a) Electron DOS for varying number $\{7, 13,\ 25,\ 37\}$ of scattering points within the \emph{vacancy} site. Faint line shows the unperturbed electron DOS for reference. (b) Perturbed divided by unperturbed electron DOS. Inset shows the geometry of the modified lattice including to the vacancy site. (c) and (d) show the spatial spectral density for the cases of (c) 7 and (d) 13 scattering points at the vacancy site. (e) Spatial spectral density for single impurity in the plaquette position.}
\label{fig4}
\end{figure}

Next, we consider impurity scattering in molecular graphene. Before we proceed, however, we briefly introduce the salient features predicted for impurity scattering in graphene. It has been shown that simple potential scattering at the position $\bfR_0$ gives rise to a resonance in the local electron DOS below the Dirac point but within the linear part of the spectrum.\cite{wehling2007,jafri2010,carva2010} We discuss the results within a nearest neighbor interaction model for graphene of the type
\begin{align}
\Hamil=&
	-t\sum_{\av{mn}\sigma}\Psi^\dagger_{m\sigma}\sigma^x\Psi_{n\sigma}
	+\sum_{m\sigma}\Psi^\dagger_{m\sigma}\bfU(\bfr_m)\Psi_{m\sigma},
\end{align}
where $\Psi_{m\sigma}^\dagger=(a_{m\sigma}^\dagger\ b_{m\sigma}^\dagger)$ denotes the pseudo-spinor for creation of electrons in the $A$- and $B$-sublattice, respectively, $t$ is the hopping parameter, $\sigma^x$ is the $x$-component of the Pauli matrices. In the last term $\bfU(\bfr_m)=\diag{U_A(\bfr_m)\ U_B(\bfr_m)}{}$ is a diagonal matrix representing the scattering potential and its coupling to the $A$- and $B$-sublattice, respectively. In this form we can describe scattering from a single vacancy in graphene by letting for example $\bfU(\bfR_0)=\diag{U_A(\bfR_0)\ 0}{}$, where $\bfR_0$ is a C site in the $A$-sublattice. This describes scattering off a potential $U_A$ in the $A$-sublattice but none in the $B$-sublattice.

Calculating the resulting electronic structure due to this potential scattering through, for example, a $T$-matrix approach, see Refs. \onlinecite{jafri2010,carva2010,wehling2007} for more details, one can write the real space local electron DOS in the $B$-sublattice as $N_B(\bfr,\omega)=\omega N_0+\delta N_B(\bfr,\omega)$, where
\begin{align}
\delta N_B(\bfr,\omega)=&
	-N_0J_0^2(k_F|\bfr-\bfR_0|)
		\im\frac{(2D+i\pi)^2}{U_A^{-1}-\omega[2\ln(D/|\omega|)+i\pi]}.
\label{eq-mgs}
\end{align}
Here, $N_0$ is related to the Fermi velocity $v_F$ and a high-energy cut-off $D\sim5 - 10$ eV,\cite{jafri2010} $k_F$ is the Fermi wave vector, and $J_0(x)$ is a Bessel function of the first kind. The correction $\delta N_B(\bfr,\omega)$ displays a clear divergent characteristics for $2\omega\ln(D/|\omega|)\rightarrow1/U_A$ in a spatial neighborhood around the vacancy. Hence, the potential scattering creates a resonance within the linear part of the spectrum for large scattering potentials $U_A$. This feature is tightly associated with the density in the $B$-sublattice, which can be seen since the corresponding electron DOS in the $A$-sublattice acquires the form $N_A(\bfr,\omega)=N_0\omega[1+J_0^2(k_F|\bfr-\bfR_0|)]$ (for large $U_A$). It is clear that the modified density in the $A$-sublattice lacks the divergent component. A three-fold spatial symmetry is therefore expected to emerge for the electron density around the vacancy.

In molecular graphene, the vacancy defect is realized by adding a scattering center, henceforth referred to as the \emph{vacancy site}, at a point corresponding to a C site in, say, the $A$-sublattice, see inset of Fig. \ref{fig4} (b). The vacancy site, breaks the \emph{bond} between this C site and the adjacent C sites in the $B$-sublattice which locally depletes the electron density at this site, in agreement with the expected features around a vacancy site. In Fig. \ref{fig4} (a) we plot the local electron DOS in the lattice sites adjacent to the vacancy site for increasing potential strength. The potential strength is modulated by varying the number of scattering points comprised in the vacancy site. The plots illustrate how the resonance builds up within the linear part of the spectrum for increasing number of scattering points in the vacancy site, in agreement with Eq. (\ref{eq-mgs}), from a minor hump for the weakest scattering potential (solid) to a more pronounced shoulder (dashed, dash-dotted, dotted) for the stronger ones. By dividing out the unperturbed electron DOS, Fig. \ref{fig4} (a) (faint), from the perturbed ones, those features become more apparent, see Fig. \ref{fig4} (b). The plots illustrate the rise of a rather sharp resonance for the strong scattering potentials. Also the topography around the scattering potential is modulated by the increased potential strength, see Fig. \ref{fig4} (c) and (d), which show the spectral density for (c) 7 and (d) 13 scattering points within the vacancy site. As expected, the vacancy scattering generates a three fold symmetric spatial signature in the $B$-sublattice, as well as an increased electron density with increasing potential strength. The emergence of the resonance within the linear part of the spectrum, sometimes referred to as a midgap state, is caused by breaking of the sublattice symmetry. The relatively weak resonance obtained in our computations compared to the prediction made in terms of Eq. (\ref{eq-mgs}) is reasonable since our finite lattice does not generate the very strong electronic confinement as acquired in proper a two-dimensional structure. Moreover, the effective potential at the vacancy site is most likely still in the weak limit, even in the case of 37 scattering points in the vacancy site. Our computations, nevertheless, demonstrate the correct tendency of the resulting electronic density around the vacancy.

Scattering off an impurity at a plaquette position (inside a hexagon) can be accounted for by removing one of the scattering sites in the triangular lattice, hence, allowing electron density to fill the void and create a bond across the hexagon through the impurity, see the spectral density plotted in Fig. \ref{fig4} (e). The bright spot signifies the increased electron density at the impurity site, and the associated six-fold symmetry of the spatial signature surrounding it is apparent. In this case a resonance builds up in both sublattices for strong enough scattering potential since the impurity couples equally strong to both.

In summary, we have studied molecular graphene using a scattering theoretical approach. Molecular graphene is constructed from the redistribution of the electron density constrained by a triangular lattice of scattering centers, dual to the honeycomb lattice. Making use of point like scattering potentials, we find that scattering centers comprising several (about 7) scattering points distributed in a hexagonal shape with diameter between roughly 3 and 6 \AA\ is sufficient to obtain a V-shaped spectrum. This shows that spatially continuous scattering potentials are not necessary to achieve realistic electronic structures. Combining those scattering centers with a moderately sized finite lattice ($\sim 13\times14$ scattering centers), we reproduce the results in Ref. \onlinecite{gomes2012} in very good agreement.

Impurity scattering can be realized by adding or removing scattering centers in the system. An important finding is that we verify that local defects may give rise to sharp resonances within the V-shaped DOS, provided that the scattering potential is sufficiently strong. This is in good agreement with previous theoretical predictions.\cite{wehling2007,jafri2010,carva2010} Our results suggest that defects of the same species as the surrounding lattice does comprise a scattering potential which is sufficiently strong to generate a resonance in the local DOS, which we believe to be the cause for the lack of experimental confirmation of this feature. Entities that correspond to stronger scattering potentials which, thereby, deplete the electron density more efficiently than the surrounding lattice should be accessible to the present experimental state-of-the-art. Thus, experimental verification of the theoretical prediction should be within the realms of near future experiments.

\section*{Acknowledgements}
We thank A. V. Balatsky and O. Eriksson for useful conversations. Support from the Swedish Research Council is acknowledged.


\begin{thebibliography}{20}
\bibitem{novoselov2004} K. S. Novoselov, A. K. Geim, S. V. Morozov, D. Jiang, Y. Zhang, S. V. Dubonos, I. V. Grigorieva, A. A. Firsov, Science, {\bf 306}, 666 (2004).
\bibitem{geim2007} A. K. Geim and K. S. Novoselov, Nat. Mater. {\bf 6}, 183 (2007).
\bibitem{katsnelson2007} M. I. Katsnelson, Mater. Today, {\bf 10}, 20 (2007).
\bibitem{castroneto2009} A. H. Castro Neto, F. Guinea, N. M. R. Peres, K. S. Novoselov, and
A. K. Geim, Rev. Mod. Phys. {\bf 81}, 109 (2009).
\bibitem{vozmediano2010} M. A. H. Vozmediano, M. I. Katsnelson, and F. Guinea, Phys. Rep.
{\bf 496}, 109 (2010).

\bibitem{elias2009} D. C. Elias, R. R. Nair, T. M. G. Mohiuddin, S. V. Morozov, P. Blake, M. P. Halsall, A. C. Ferrari, D. W. Boukhvalov, M. I. Katsnelson, A. K. Geim, K. S. Novoselov, Science, {\bf 323}, 610 (2009).

\bibitem{polini2013} M. Polini, F. Guinea, M. Lewenstein, H. C. Manoharan, and V. Pellegrini, Nat. Nanotech. {\bf 8}, 625 (2013).

\bibitem{peleg2007} O. Peleg, G. Bartal, B. Freedman, O. Manela, M. Segev, and D. N. Christodoulides, Phys. Rev. Lett. {\bf 98}, 103901 (2007).
\bibitem{kuhl2010} U. Kuhl, S. Barkhofen, T. Tudorovskiy, H.-J. St\"ockmann, T. Hossain, L. de Forges de Parny, and F. Mortessagne, Phys. Rev. B {\bf 82}, 094308 (2010).
\bibitem{singha2011} A. Singha, M. Gibertini, B. Karmakar, S. Yuan, M. Polini, G. Vignale, M. I. Katsnelson, A. Pinczuk, L. N. Pfeiffer, K. W. West, and V. Pellegrini, Science, {\bf 332}, 1176 (2011).
\bibitem{gomes2012} K. K. Gomes, W. Mar, W. Ko, F. Guinea, and H. C. Manoharan, Nature (London), {\bf 483}, 306 (2012).
\bibitem{soltan2011} P. Soltan-Panahi, J. Struck, P. Hauke, A. Bick, W. Plenkers, G. Meineke, C. Becker, P. Windpassinger, M. Lewenstein, and K. Sengstock, Nature Physics, {\bf 7}, 434 (2011).
\bibitem{tarruell2012} L. Tarruell, D. Greif, T. Uehlinger, G. Jotzu, and T. Esslinger, Nature (London), {\bf 483}, 302 (2012).

\bibitem{haldane1988} F. D. M. Haldane, Phys. Rev. Lett. {\bf 61}, 2015 (1988).
\bibitem{kane2005} C. L. Kane and E. J. Mele, Phys. Rev. Lett. {\bf 95}, 226801 (2005).
\bibitem{guinea2010} F. Guinea, M. I. Katsnelson, and A. K. Geim, Nature Physics, {\bf 6}, 30 (2010).
\bibitem{meng2010} Z. Y. Meng, T. C. Lang, S. Wessel, F. F. Assaad, and A. Muramatsu, Nature (London), {\bf 464}, 847 (2010).

\bibitem{fiete2003} G. A. Fiete and E. J. Heller, Rev. Mod. Phys. {\bf 75}, 933 (2003).

\bibitem{wunsch2008} B. Wunsch, F. Guinea, and F. Sols, New J. Phys. {\bf 10}, 103027 (2008).
\bibitem{park2009} C. -H. Park and S. G. Louie, Nano. Lett. {\bf 9}, 1793 (2009).
\bibitem{gibertini2009} M. Gibertini, A. Singha, V. Pellegrini, M. Polini, G. Vignale, A. Pinczuk, L. N. Pfeiffer, and K. W. West, Phys. Rev. B, {\bf 79}, 241406 (2009).

\bibitem{fransson2010} J. Fransson, H. C. Manoharan, and A. V. Balatsky, Nano Lett. {\bf 10}, 1600 (2010).
\bibitem{gawronski2011} H. Gawronski, J. Fransson, and K. Morgenstern, Nano Lett. {\bf 11}, 2720 (2011).
\bibitem{she2013} J. -H. She, J. Fransson, A. R. Bishop, and A. V. Balatsky, Phys. Rev. Lett. {\bf 110}, 026802 (2013).
\bibitem{fransson2013} J. Fransson, J. -H. She, L. Pietronero, and A. V. Balatsky, Phys. Rev. B, {\bf 87}, 245404 (2013).

\bibitem{hirjibehedin2006} C. F. Hirjibehedin, C. P. Lutz, A. J. Heinrich, Science, {\bf 312}, 1021 (2006).
\bibitem{zhou2010} L.   Zhou, J. Wiebe, S. Lounis, E. Vedmedenko, F. Meier, S. Bl\"ugel, P. H. Dederichs, and R. Wiesendanger, Nat. Phys. {\bf 6}, 187 (2010).
\bibitem{decker2011} R. Decker, Y. Wang, V. W. Brar, W. Regan, H. -Z. Tsai, Q. Wu, W. Gannett, A. Zettl, and M. F. Crommie, Nano Lett. {\bf 11}, 2291 (2011).
\bibitem{wang2012} Y. Wang, V. W. Brar, A. V. Shytov, Q. Wu, W. Regan, H. -Z. Tsai, A. Zettl, L. S. Levitov, and M. F. Crommie, Nat. Phys. {\bf 8}, 653 (2012).

\bibitem{khajetoorians2011} A. A. Khajetoorians, J. Wiebe, B. Chilian, R. Wiesendanger, Science, {\bf 332}, 1062 (2011). 
\bibitem{loth2012} S. Loth, S. Baumann, C. P. Lutz, D. M. Eigler, A. J. Heinrich, Science, {\bf 335}, 196 (2012).
\bibitem{moon2008} C. R. Moon, L. S. Mattos, B. K. Foster, G. Zeltzer, W. Ko, H. C. Manoharan, Science, {\bf 319}, 782 (2008).

\bibitem{binnig1982} G. Binnig, H. Rohrer, Ch. Gerber, and E. Weibel, Appl. Phys. Lett. {\bf 40}, 178 (1982).
\bibitem{tersoff1983} J. Tersoff and D. R. Hamann, Phys. Rev. Lett. {\bf 50}, 1998 (1983).

\bibitem{fransson2012} J. Fransson and A. V. Balatsky, Phys. Rev. B, {\bf 85}, 161401(R) (2012).

\bibitem{uehlinger2013} T. Uehlinger, G. Jotzu, M. Messer, D. Greif, W. Hofstetter, U. Bissbort, and T. Esslinger, Phys. Rev. Lett. {\bf 111}, 185307 (2013).

\bibitem{jafri2010} S. H. M. Jafri, K. Carva, E. Widenkvist, T. Blom, B. Sanyal, J. Fransson, O. Eriksson, U. Jansson, H. Grennberg, O. Karis, R. A. Quinlan, B. C. Holloway, and K. Leifer, J. Phys. D: Appl. Phys. {\bf 43}, 045404 (2010).
\bibitem{carva2010} K. Carva, B. Sanyal, J. Fransson, and O. Eriksson, Phys. Rev. B {\bf 81}, 245405 (2010).

\bibitem{wehling2007}  T. O. Wehling, A. V. Balatsky, M. I. Katsnelson, A. I. Lichtenstein, K. Scharnberg, and R. Wiesendanger, Phys. Rev. B {\bf 75}, 125425 (2007).


\end{thebibliography}
\end{document}